\documentstyle[aps,12pt]{revtex}
\begin{document}

\title{TOMOGRAPHY OF SOLITONS \footnote{Presented by M.A. Man'ko
at the International Workshop``Nonlinear Physics: Theory
and Experiment. II'' (Gallipoli, Lecce, Italy, 27 June --
6 July 2002)}}
\author{{\bf Sergio De Nicola},$^1$, {\bf Renato Fedele},$^2$
{\bf Margarita A. Man'ko}$^3$ and {\bf Vladimir I. Man'ko}$^3$}

\medskip
\address{$1$ Istituto di Cibernetica "Eduardo Caianiello" del CNR
Comprensorio "A. Olivetti", Fabbr. 70\\ Via Campi  Flegrei, 34,  I-80078
Pozzuoli (NA), Italy\\ E-mail:  s.denicola@cib.na.cnr.it
\\
\medskip
$^2$ Dipartimento di Scienze Fisiche, Universit\`{a} Federico II and INFN,
\\ Complesso Universitario di M.S. Angelo,\\ Via Cintia, I-80126 Napoli,
Italy\\ Email: renato.fedele@na.infn.it\\
\medskip
$^3$ P. N. Lebedev Physical Institute,\\ 119991 Moscow, Russia\\ Email(s):
mmanko@sci.lebedev.ru, manko@sci.lebedev.ru}

\maketitle

\begin{center}
{\bf Abstract}
\end{center}
\noindent
{We develope the tomographic representation of
wavefunctions which are solutions of the generalized
nonlinear Schr\"{o}dinger equation (NLSE) and show its
connection with the Weyl--Wigner map. The generalized
NLSE is presented in the form of a nonlinear
Fokker--Planck--type equation for the standard
probability distribution function (marginal
distribution). In particular, this theory is applied to
the envelope solitons, where tomograms for envelope
bright solitons of a wide family of modified NLSE are
presented and numerically evalueted.}

\bigskip
\noindent
{\small PACS number(s):
\noindent
52.35.Mw - Nonlinear waves;
\noindent
05.45.Yv - Solitons;
\noindent
42.65 - Nonlinear optics;
\noindent
67.57.Jj
- Collective modes in quantum fluids}

\medskip
\begin{center}
{\bf Figures available at the web site:\\
http://people.na.infn.it/$\sim$fedele/tomosol-figs/
tomosol-fig.htm}
\end{center}
\newpage

\section{Introduction}
In quantum mechanics and quantum optics
as well as in signal analysis, the properties of wave
functions satisfying the linear Schr\"{o}dinger equation and the properties
of complex analytic signals were studied using the tomographic probability
distribution functions called tomograms.
In the tomographic representation,
the complex wave function is associated with the standard probability
distribution by means of invertible map.
In signal processing, this method was called noncommutative tomography
of analytic signal.

Several tomographic schemes are known, i.e., optical tomography,
symplectic tomography, photon-number tomography, and spin tomography.
An advantage to use the tomographic representation for quantum states is
related to the possibility of measuring the states, described in the
standard approach by complex wave functions.
The tomographic map of the measured tomogram provides the possibility
to reconstruct the wave function (up to the constant phase factor) of
the quantum state.

Till now the tomographic representation was used for discussing the solutions
of linear Schr\"{o}dinger equation of quantum mechanics.

On the other hand, there exist nonlinear processes
described by well-known nonlinear equations like
nonlinear Schr\"{o}dinger equation with cubic nonlinearity or
more complicated nonlinearities. There exist a huge
literature to construct solutions of nonlinear equations
and different methods to approach the problem. Among
solutions of nonlinear equations, there is a set of
specific soliton solutions.

Up to our knowledge, till now the tomographic
representation was not used to describe the soliton
solutions of nonlinear equations. We introduce here the
tomographic representation of the soliton solutions of
nonlinear equations and concentrate on tomograms of
envelope solitons of nonlinear Schr\"{o}dinger equation.

\section{Weyl--Ville--Wigner map}
Now we review the approach called symplectic tomography
of quantum state. The real meaning of this scheme is the
map of the complex wave function $\psi(x)$ of a real
variable $x$ $(-\infty<x<\infty)$ onto a family of
probability distributions $w(X,\mu,\nu)$ of a random real
variable $X$ $(-\infty<X<\infty)$ labeled by two real
parameters $\mu$ $(-\infty<\mu<\infty)$ and $\nu$
$(-\infty<\nu<\infty)$.

We use dimensionless variables. The map may be realized
by the following steps. First, one constructs the density
matrix~\cite{1}
which is a complex function of two variables $x$ and $x'$
\begin{equation}\label{1}
\rho_\psi(x,x')=\psi(x)\psi^*(x').
\end{equation}
Then one uses Wigner--Weyl map of density matrix onto
real Wigner function on phase space $W(q,p)$~\cite{2}
 of two real variables $p$ and $q$
\begin{equation}\label{2}
W_\psi(q,p)=\int\rho_\psi\Big(q+{u}/{2}\,,
q-{u}/{2}\Big)e^{-ipu}\,du.
\end{equation}
The Wigner function takes real values. If the wave
function is normalized
$$\int |\psi(x)|^2\,dx=1,$$
the Wigner function is also normalized
$$\int W_\psi(q,p)\,{dq\,dp}/{2\pi}=1.$$
Ville~\cite{3} used this map in the analytic signal
theory.


The inverse of the Fourier transform (\ref{2}) defining
the Wigner
 function in terms of density matrix reads
\begin{equation}\label{5}
\psi(x)\psi^*(x')=\frac{1}{2\pi}\int W_\psi\left(\frac{x+x'}{2}\,,p\right)
e^{ip(x-x')}\,dp.
\end{equation}
One can see that the Wigner function determines the complex wave
function up to the constant factor
\begin{equation}\label{6}
\psi^*(0)\psi(x)=\frac{1}{2\pi}\int W_\psi\left(\frac{x}{2}\,,p\right)
e^{ipx}\,dp.
\end{equation}
Modulus of the constant factor $|\psi(0)|$
is determined by the relationship
\begin{equation}\label{7}
|\psi(0)|^2=\frac{1}{2\pi}\int W_\psi\left(0,p\right)\,dp.
\end{equation}
We suppose that $\psi(0)$ is not equal to zero. Thus,
given Wigner function one can reconstruct the complex
wave function up to constant phase factor. This means
that the Wigner function contains the same information
which the density matrix does. Also this means that the
Wigner function contains the same information as the wave
function $\psi(x)$ does (up to the constant phase
factor). The Wigner function can be identified with
so-called Weyl symbol of density operator describing the
quantum state. There exist diferent kinds of symbols of
operators. As it was shown recently~\cite{4} tomograms
can be also identified with a specific symbol of a
density operator.

\section{Tomographic map}
Let us now construct the tomographic map. To do this, we use the integral
Radon transform of the Wigner function
\begin{equation}\label{8}
w(X,\mu,\nu)=\int W_\psi(q,p)\delta(X-\mu q-\nu p)\,\frac{dq\,dp}{2\pi}\,.
\end{equation}
In this formula, the Dirac delta-function term
$\delta(X-\mu q-\nu p)$ collects values of the Wigner
function $W_\psi(q,p)$ from the line in the phase space
which is described by the expression obtained by equating
the argument of Dirac delta-function to zero. One can
prove~\cite{5} that for normalized Wigner function the
function $w(X,\mu,\nu)$ is normalized probability
distribution function of random variable $X$ (called
tomogram of the Wigner function), i.e., one has
\begin{equation}\label{9}
\int w(X,\mu,\nu)\,dX=1.
\end{equation}
This tomogram was called symplectic tomogram~\cite{6},
since it is related to linear symplectic transform in the phase space.

There is another form of the map~(\ref{8}) given in terms of Fourier transform
(we used Fourier transform of delta-term in (\ref{8}))
\begin{equation}\label{10}
w(X,\mu,\nu)=\int W_\psi(q,p)\exp\left[ik(X-\mu q-\nu p)\right]\,
\frac{dk\,dq\,dp}{(2\pi)^2}.
\end{equation}
The Fourier integral form of the tomogram (\ref{10})
gives the possibility to get easily the inverse
transform~\cite{7}
\begin{equation}\label{11}
W(q,p)=\int w(X,\mu, \nu)\exp\left[i(X-\mu q-\nu p)\right]\,
\frac{dX\,d\mu\,d\nu}{2\pi}\,.
\end{equation}
Thus, given the symplectic tomogram $w(X,\mu,\nu)$ one can find
the Wigner function $W(q,p)$. Using known relationships (\ref{1}), (\ref{2})
and (\ref{5}) of the Wigner function $W(q,p)$ and the wave function $\psi(x)$,
 one can obtain the expression for tomograms in terms of the wave function~\cite{8}
\begin{equation}\label{12}
w(X,\mu,\nu)=\frac{1}{2\pi|\nu|}\left|\int \psi(y)
\exp\left(\frac{i\mu}{2\nu}\,y^2-i\,\frac{Xy}{\nu}\right)\,dy\right|^2.
\end{equation}
This formula shows that for a complex function $\psi(x)$
one can find the tomogram. Formula~(\ref{12}) was found
for the quantum wave function and for noncommutative
tomography of the analytic signal~\cite{8}.
But it can be used for other arbitrary aims as well. The goal which we are
 going to reach is to use this formula for the description of solitons.

\section{Tomogram as Fourier transform of a chirped soliton}
There are some properties of tomograms to be used. The
homogeneity property~\cite{9}
follows from relations (\ref{8}) and (\ref{12})
\begin{equation}\label{14a}
w(\lambda X,\lambda \mu, \lambda\nu)=\frac{1}{|\lambda|}w(X,\mu,\nu).
\end{equation}
This means that, in reality, the tomogram is the function of two real
variables. For example, one can take
\begin{equation}\label{14b}
\mu=\cos\theta,\qquad \nu=\sin\theta.
\end{equation}
In this case, the symplectic tomogram is given by the formula
\begin{equation}\label{15}
w(X,\theta)=\int W(q,p)\delta(X-q\cos\theta -p\sin\theta)\frac{dq\,dp}{2\pi}.
\end{equation}
This relation of the Wigner function $W(q,p)$ and the
tomographic probability $w(X,\theta)$ (called tomogram of
the optical tomography scheme) was used in the signal
theory in \cite{10}. The relation of the optical tomogram
$w(X,\theta)$ to the wave function reads
 \begin{equation}\label{16}
w(X,\theta)=\frac{1}{2\pi|\sin\theta|}\left|\int\psi(y)
\exp\left(\frac{i}{2}\,\cot\theta \,y^2-\frac{iXy}{\sin\theta}
\right)dy\right|^2.
\end{equation}
The integrand in the above expression (\ref{16}) is
similar to the Green function of the quantum harmonic
oscillator and the tomogram coincides with modulus
squared of fractional Fourier transform of
$\psi(y)$~\cite{11}.
The tomogram $w(X,\theta)$ is used in quantum optics in
the scheme of measuring quantum photon states by means of
the so-called optical homodyne tomography~\cite{12}.
It was also discussed in the context of quantum-optics
measurements in \cite{13}.
Thus the tomogram $w(X,\mu,\nu)$ determines completely
the Wigner function and the complex function $\psi(x)$ (up to the phase factor),
if it is known for real parameters satisfying the constraint
\begin{equation}\label{17}
\mu^2+\nu^2=1.
\end{equation}
But one can use other constraints too. Thus, the
homogeneity property (\ref{14a}) implies that the
particular values of the symplectic tomogram, e.g.,
$w(1,\mu,\nu)$, $w(X,1,\nu)$ and $w(X,\mu,1)$ determine
the whole tomogram and, consequently, the complex
function $\psi(x)$ (up to the phase factor) and the
Wigner function $W(q,p)$ completely.

Another property can be obtained by change of variables in (\ref{12})
\begin{equation}\label{19}
\frac{y}{\nu}=z,
\end{equation}
which gives the following expression for the tomogram
\begin{equation}\label{20}
w(X,\mu,\nu)=\frac{|\nu|}{2\pi}\left|\int\widetilde\psi(z,\mu,\nu)
e^{-iXz}dz\right|^2,
\end{equation}
where the function $\widetilde\psi(z,\mu,\nu)$ is the function describing
``the chirped soliton'' (we mean that $\psi(y)$ in (\ref{12}) is considered
as a soliton solution of a nonlinear equation)
\begin{equation}\label{21}
\widetilde\psi(z,\mu,\nu)=\psi(z\nu)\exp\left(\frac{i}{2}\,\mu\nu z^2\right).
\end{equation}

Expression (\ref{20}) is convenient for numerical calculations because it gives
the tomogram in terms of the standard Fourier transform of the chirped soliton.
In terms of optical tomogram, the expression can be rewritten as
\begin{equation}\label{22}
w(X,\theta)=\frac{|\sin\theta|}{2\pi}\left|\int\widetilde\psi(z,\theta)
e^{-iXz}dz\right|^2,
\end{equation}
where the chirped soliton has the form
\begin{equation}\label{23}
\widetilde\psi(z,\theta)=\psi(z\sin\theta)\exp\left(\frac{iz^2}{4}\,
\sin 2\theta\right).
\end{equation}
The formula obtained is used below to make plots of
solitons.

\section{Nonlinear equations in tomographic and Weyl--Wigner--Moyal
representations}
Let us consider the following
generalized nonlinear Schr\"{o}dinger equation (NLSE):
\begin{equation}
i{\partial\psi\over \partial s} = -{1\over
2}{\partial^2\psi\over\partial x^2} +
U\left[|\psi|^2\right]\psi,
\label{gnlse}
\end{equation}
where $s$ and $x$ are the time-like and space-like
variables and $\psi = \psi (x,s)$ is a complex wave
function describing the system's evolution in the
configuration space; $U=U\left[|\psi|^2\right]$ is an
arbitrary real functional of $|\psi|^2$.

In this section, we derive the evolution equations for
solitons in the phase-space representation. For the case
of the cubic NLSE
\begin{equation}\label{13}
i\frac{\partial \psi}{\partial
s}=-\frac{1}{2}\,\frac{\partial ^2\psi}{\partial
x^2}+q_0|\psi|^2\psi,
\end{equation}
the density matrix (\ref{1}) satisfies the following evolution equation:
\begin{eqnarray}\label{24}
&&i\frac{\partial\rho(x,x',s)}{\partial
s}=-\frac{1}{2}\left(
\frac{\partial^2}{\partial x^2}-\frac{\partial^2}{\partial x'^2}\right)
\rho(x,x',s)+ \nonumber\\
&&+\left\{V\left[\int\delta(x-y)\rho(x,y,s)\,dy\right]-V
\left[\int\delta(x'-y)\rho(x',y,s)\,dy\right]\right\}\rho(x,x',s),
\end{eqnarray}
where
\begin{equation}\label{25}
V\left[\int\delta(x-y)\rho(x,y,s)\,dy\right]=q_0\int
\delta(x-y)\rho(x,y,s)\,dy
\end{equation}
is the potential-energy functional of the density matrix
for nonlinear Schr\"{o}dinger equation~(\ref{13}) with cubic
nonlinearity.

The transition to the evolution equation for the Wigner function can be done
using the standard algebra, which provides the following recepie.
One has to make in (\ref{24}) the replacement $\rho\rightarrow W$ along with
the following replacement:
\begin{eqnarray}\label{eso29s}
\frac {\partial }{\partial x}\,\rho \,(x,\,x^\prime )&\longrightarrow &
\left (\frac {1}{2}\,\frac {\partial }{\partial q}+i\,p\right )
\,W\,(q,\,p)\,;\nonumber\\
\frac {\partial }{\partial x^\prime }\,\rho \,(x,\,x^\prime )&\longrightarrow &
\left (\frac {1}{2}\,\frac {\partial }{\partial q}-i\,p\right )
\,W\,(q,\,p)\,;\nonumber\\[-2mm]
&&\\[-2mm]
x\,\rho \,(x,\,x^\prime )&\longrightarrow &
\left (q+\frac {i}{2}\,\frac {\partial }{\partial p}\right )
\,W\,(q,\,p)\,;\nonumber\\
x^\prime \,\rho \,(x,\,x^\prime )&\longrightarrow &
\left (q-\frac {i}{2}\,\frac {\partial }{\partial p}\right )
\,W\,(q,\,p)\,.\nonumber
\end{eqnarray}
It provides the following Moyal-like form of the nonlinear equation (\ref{24})
for the Wigner function
\begin{equation}\label{26}
\frac{\partial W(q,p,s)}{\partial s}=-
p\frac{\partial W(q,p,s)}{\partial q}+\frac{1}{i}
\left\{V\left[\rho\left(q+\frac{i}{2}\frac{\partial}{\partial p}\,,
q+\frac{i}{2}\frac{\partial}{\partial p}\,, s\right)\right]
-\mbox{c.c.}\right\}W(q,p,s).
\end{equation}
Here the arguments of the potential energy are replaced
by the operators which act onto the Wigner function. For
the cubic NLSE
\begin{equation}\label{27}
V(z)=q_0z,
\end{equation}
one has a simple form of the equation
\begin{equation}\label{28}
\frac{\partial W(q,p,s)}{\partial s}+p\,
\frac{\partial}{\partial q}W(q,p,s)
-2q_0\mbox{Im}\,\rho\left(q+\frac{i}{2}\frac{\partial}{\partial
p}\,, q+\frac{i}{2}\frac{\partial}{\partial p}\,,
s\right)W(q,p,s)=0.
\end{equation}

Using the relation
\begin{equation}\label{29}
\rho(x,x)=\int W(x,p)\,\frac{dp}{2\pi}\,,
\end{equation}
one has
\begin{equation}\label{30}
\frac{\partial W(q,p,s)}{\partial s}+p\,
\frac{\partial W(q,p,s)}{\partial q}
-2q_0\,\mbox{Im}\,\int
W\left(q+\frac{i}{2}\frac{\partial}{\partial p}\,,
P,s\right)\frac{dP}{2\pi}\,W(q,p,s)=0.
\end{equation}
In (\ref{26}), (\ref{28}), and (\ref{30}) the arguments
of the density matrix and Wigner function are replaced by
operators and the operators act on the Wigner function
itself. Equations (\ref{26}) and (\ref{28}) can be
presented in the form of Moyal-like series~\cite{14}.
Equation (\ref{26}) for an arbitrary nonlinear potential $V(z)$ can be
written in terms of the functional partial differential equation for
the Wigner function only
\begin{equation}\label{31}
\frac{\partial W(q,p,s)}{\partial s}+p\,
\frac{\partial W(q,p,s)}{\partial q}
-2\mbox{Im}\left\{V\left[\int W\left(q+\frac{i}{2}
\frac{\partial}{\partial p}\,,
P,s\right)\,\frac{dP}{2\pi}\right]\right\}W(q,p,s)=0.
\end{equation}
Now we consider the relation of the density matrix, Wigner function and tomogram
\begin{equation}
W(q,p)=\frac {1}{2\pi}\int w(X,\mu,\nu )\exp \,[-i (\mu q
+\nu p-X)]\,d\mu \,d\nu \,dX~~~;
\label{tila}
\end{equation}
\begin{equation}
\rho (X,X^\prime )=\frac {1}{2\pi}
\int w(Y,\mu ,X-X^\prime )
\exp\left [i
\left (Y-\mu \,\frac {X+X^\prime }{2}\right )\right ]\,d\mu \,dY.
\label{tilb}
\end{equation}
In view of this relation, the following rule for
substitution in the evolution equation (\ref{24})
\begin{eqnarray}\label{32}
&&\rho(x,x',s)\longrightarrow w(X,\mu,\nu,s);\nonumber\\
&&x\rho \longrightarrow
\left[-\left(\frac {\partial }{\partial X}\right)^{-1}
\frac {\partial }{\partial \mu}
+\frac{i}{2}\,\nu\,\frac{\partial}{\partial X}\right]w;\nonumber\\
&&\frac{\partial}{\partial x}\rho \longrightarrow \left[\frac{\mu}{2}
\,\frac {\partial }{\partial X}
-i\left(\frac {\partial }{\partial X}\right)^{-1}
\frac{\partial}{\partial \nu}\right]w;\\
&&x'\rho\longrightarrow \left[
-\left(\frac {\partial }{\partial X}\right)^{-1}
\frac {\partial }{\partial \mu}
-\frac{i}{2}\,\nu\,\frac{\partial}{\partial X}\right]w;
\nonumber\\
&&\frac{\partial}{\partial x'}\rho\longrightarrow \left[\frac{\mu}{2}
\,\frac {\partial }{\partial X}
+i\left(\frac {\partial }{\partial X}\right)^{-1}
\frac{\partial}{\partial \nu}\right]w
\nonumber
\end{eqnarray}
provides the tomographic form of the nonlinear equation
under consideration
\begin{eqnarray}\label{33}
&&\frac{\partial w(X,\mu,\nu,s)}{\partial s}
+\mu\,\frac{\partial w(X,\mu,\nu,s)}{\partial
\nu} + \nonumber\\ &&-2\,\mbox{Im}\,V\left\{\int
w(y,\mu',0,s)\exp\left[i\left(y+\mu'\left[
\left(\frac{\partial}{\partial X}\right)^{-1}\frac{\partial}{\partial \mu}
-\frac{i}{2}\,\nu\,\frac{\partial}{\partial X}\right]\right)\right]
\frac{dy\,d\mu'}{2\pi}\right\}w(X,\mu,\nu,s)=0.
\end{eqnarray}
In the above equation~(\ref{33}), one has the integro-differential operator
 in the exponent, which acts on the tomographic probability function.
The integral operator $\left(\partial/\partial X\right)^{-1}$ is defined
by the following action on Fourier component of the function $f(X)$:
$$
\left(\frac{\partial}{\partial X}\right)^{-1}f(X)=
\left(\frac{\partial}{\partial X}\right)^{-1}\int\widetilde f(k)e^{ikX}dk=
\int\frac{\widetilde f(k)}{ik}\,e^{ikX}dk.
$$

For the case of the cubic NLSE, one has
\begin{eqnarray}\label{34}
&&\frac{\partial w(X,\mu,\nu,s)}{\partial s}
+\mu\,\frac{\partial w(X,\mu,\nu,s)}{\partial
\nu} + \nonumber\\&&-2q_0\,\mbox{Im}\left\{\int
w(y,\mu',0,s)\exp\left[i\left(y+\mu'\left[
\left(\frac{\partial}{\partial X}\right)^{-1}\frac{\partial}{\partial \mu}
-\frac{i}{2}\,\nu\,\frac{\partial}{\partial X}\right]\right)\right]
\frac{dy\,d\mu'}{2\pi}\right\}w(X,\mu,\nu,s)=0.
\end{eqnarray}
It should be pointed out that Eq. (\ref{34}) has the solutions which
in the course of the evolution process preserve the positivity and
normalization. Thus the soliton solutions of the nonlinear equations
can be mapped onto probability distribution functions.

The meaning of the probability distributions is the following.
If $x$ is a coordinate and $p$ is the momentum, the value
$X=\mu x +\nu p$
is the position in the reference frame in the phase space $(x,p)$,
the reference frame being scaled and rotated. The parameters of the
scaling $\lambda$ and rotation $\theta$ are determined by the real
parameters $\mu$ and $\nu$, namely,
$\mu=e^\lambda\cos\theta$ and $\nu=e^{-\lambda}\sin\theta.$

The probability distributions $w(X,\mu,\nu)$ determine
soliton solutions $\psi(x)$ in the corresponding
representation. Consequently, in the tomographic
representation soliton solutions of nonlinear dynamic
systems are solutions of generalized Fokker--Planck-type
equations for the standard probability distributions.
Such representation can be useful from mathematical point
of view since analysis of probabilities and their
assymptotics can be additionally incorporated (using
existing theorems) on the behaviour of the probability
distribution functions.

\section{Examples}
In this section, we study bright soliton in both the
tomographic and Weyl--Wigner representations. The
envelope bright soliton of the cubic NLSE, i.e.
\begin{equation}
\Psi
(x,s)=\left({\frac{2\left|E\right|}{\left|q_0\right|}}
\right)^{1/2}
\mbox{sech}\left[{\sqrt{2\left|E\right|}}~\xi\right]
\exp\left[i\left[V_0 x
-\left(E+V_{0}^{2}/2\right)s\right]\right]~~~,
\label{soliton-envelope}
\end{equation}
where $E$ is a negative real constant, $V_0$ is an
arbitrary real constant and $\xi = x-V_0 s$ (see f.i.,
Ref. \cite{17}). Thus, the corresponding optical tomogram
is given by the formula
\begin{eqnarray}\label{37}
&&w_{\rm b}(X,\theta,s)=\frac{|E\sin\theta|}{|q_0|\pi}\left|\int
\mbox{sech}\left[\sqrt{2|E|}(y\sin\theta-V_0s)
\right]\right. {\times} \nonumber\\
&&\left.{\times}\exp\left\{i\left[V_0y\sin\theta-\left(E+
\frac{V_0^2}{2}\right)s\right]+\frac{i\sin 2\theta}{4}\,y^2-iXy
\right\}dy\right|^2.\nonumber\\
&&
\end{eqnarray}
It has been recently shown that the following modified
NLSE ($U[|\psi|^2]=q_0 |\psi|^{2\beta}$)
\begin{equation}
i{\frac{\partial\Psi}{\partial s}}=-{\frac{}{2}}{\frac{
\partial^2\Psi}{\partial x^2}}+q_0|\Psi|^{2\beta}\Psi,
\label{mnlse}
\end{equation}
for $q_0 <0$ and any real positive value of $\beta$, has
the following envelope soliton-like solutions~\cite{15}:
\begin{equation}
\Psi
(x,s)=\left[{\frac{\left|E\right|\left(1+\beta\right)}
{\left|q_0\right|}}\right]^{1/2\beta}\mbox{sech}^{1/\beta}
\left[{\beta\sqrt{2\left|E\right|}}~\xi
\right]\exp\left[i\left[V_0 x
-\left(E+V_{0}^{2}/2\right)s\right]\right],
\label{mnlse-solution}
\end{equation}
where the real numbers $V_0$ and $E$ are arbitrary and
negative, respectively, and still $\xi = x-V_0 s$ (note
that $V_0$ is the soliton velocity). It should be noted
that the case $\beta =1$ (ordinary envelope bright
soliton of the cubic NLSE~\cite{16}) can be very easily
recovered~\cite{15}.

The Wigner function of bright soliton for $V_0=0$ is
given by the formula
\begin{equation}\label{38}
W(x,p)=\frac{|E|}{|q_0|}\int
\mbox{sech}\left[\sqrt{2|E|}\left(x+\frac{u}{2}
\right)\right] \mbox{sech}\left[\sqrt{2|E|}\left(x-
\frac{u}{2}\right)\right] \exp^{-ipu}du.
\end{equation}
The tomogram of the soliton solution of generalized
nonlinear Schr\"{o}dinger equation is given by the formula
$(V_0 =0)$
\begin{equation}\label{39}
w (X,\mu,\nu)=\frac{1}{2\pi|\nu|}
\left|\left[\frac{|E|(1+\beta)}{|q_o|} \right]^{1/\beta}\int
\mbox{sech}^{1/\beta}\left[\beta\sqrt{2|E|}~y\right]
\exp\left(\frac{i\mu}{2\nu}y^2
-\frac{iXy}{\nu}\right)dy\right|^2.
\end{equation}
The Wigner function of the soliton solution of
generalized nonlinear Schr\"{o}dinger equation is given by
the formula $(V_0 =0)$
\begin{equation}\label{40}
W(x,p)=\left[\frac{|E|(1+\beta)}{|q_0|}\right]^{1/\beta}\int
\mbox{sech}^{1/\beta}\left[\beta\sqrt{2|E|}
\left(x+\frac{u}{2}\right)\right] \mbox{sech}^{1/\beta}
\left[\beta\sqrt{2|E|}
\left(x-\frac{u}{2}\right)\right] \exp^{-ipu}du.
\end{equation}

\section{Some numerical computations}
3D Plots and density plots of both tomograms and Wigner
functions for bright solitons with $\beta= 0.5$, $1.0$,
$2.0$ and $2.5$ are given in Fig.1 - Fig.5. We have fixed
the free parameters as follows: $V_0 =0$, $E=-1$, and
$q_0 =-1$. Fig.1 represents the 3D plots of the tomogram
of the solitons for the different values of $\beta$. The
corresponding density plots are displayed in Fig.2~.
Fig.3 displays the 3D plots of the Wigner
quasidistribution of the solitons for the diverse values
of $\beta$. For such solitons, the phase-space regions
where the Wigner function is negative are not easily
visible in Fig.3, whilst the corresponding density plots
displayed by Fig.4 show clearly this behaviour. They
correspond to the black and grey regions. The "deepness"
of the negativity is represented in scale of grey. Black
regions correspond to the deepest negative parts. The
size of these negative parts of the Wigner function are,
for instance, clearly represented by the cross section
shown in Fig.5 at $p=2$ for the diverse values of
$\beta$.

\section{Remarks and perspectives}
We discussed the tomography of solitons considering the
tomograms as additional characteristics of the soliton
solutions of nonlinear dynamic equations. But there
exists another experimental aspect of tomograms because
the tomograms can be directly measured in different
situations. Thus the problem of tomography of some
phenomena which is described by a complex function
$\psi(x)$ is equivalent to the problem of measuring the
amplitude $|\psi(x)|$ and the phase $\varphi (x)$ of the
complex function $\psi(x)=|\psi(x)|\exp i\varphi(x)$. The
function $\psi(x)$ can describe a soliton but also this
function can describe some signal connected with
different processes, e.g., in optical fibers, in plasma,
etc. In all processes where one needs to measure the
amplitude and phase by measuring experimentally only
intensities, the tomography can be used as an instrument
for achieving this aim. We consider two different
possibilities which are based on the symplectic tomogram
(\ref{12}). The density matrix can be reconstructed
either by measuring the tomogram of the optical
tomography scheme $w(X,\theta)$ or tomogram $w(X,1,\nu)$.

Now we show that both tomograms can be obtained in two
different and realizable processes. The optical tomogram
can be rewritten in terms of fractional Fourier transform
(which is reduced to the Green function of the harmonic
oscillator)~\cite{11}.
In fact, one has
\begin{equation}\label{42}
w(X,\Theta)=\left|\frac{1}{\sqrt{2\pi i\sin\Theta}}\int\psi(y)\exp\left[
\frac{i}{2}\cot\Theta\left(y^2+X^2\right)-\frac{iXy}{\sin\Theta}\right]
\,dy\right|^2.
\end{equation}
In this formula, we take $\nu=\sin \Theta$ and $\mu=\cos
\Theta$. The phase factor $\exp iX^2/2\cot \Theta$ does
not change the value of the tomogram.

The tomogram presented in such form coincides with the value of
 the wave function at the point $x$ at the time moment $t$ if
 the initial value of the wave function at the time moment $t=0$
 is equal to $\psi(y)$. This means that to reconstruct the
initial value of the wave function $\psi(x)$ including both the
amplitude $|\psi(x)|$ and the phase $\varphi (x)$
$$\psi(x)=|\psi(x)|\exp i\varphi(x),$$ one
can measure the tomogram, i.e., amplitude squared of the
wave function which evolves in the quadratic potential
well. This situation can be perfectly done for optical
fibers with a parabolic profile of the refractive index
called ``selfoc'' (linear propagation). In fact, the
light beams in optical fibers obey to the
Schr\"{o}dinger-like equation which follows from the
Helmholtz equation in the Fock--Leontovich
approximation~\cite{18}.
But time $t$ in the Schr\"{o}dinger equation is replaced
by the longitudinal coordinate $z$ and the Planck's constant is
replaced by the wavelength. Thus, to measure the input field
amplitude and phase, it is sufficient to measure
the tomogram which is the field intensity in each
cross-section of the fiber given by
longitudinal coordinate $0< z\leq 2\pi$.

Another possibility is related to the formula
\begin{equation}\label{43}
w(X,1,\nu)=\left|\frac{1}{\sqrt{2\pi i\nu}}\int\exp\frac{i(X-y)^2}{2\nu}
\psi (y)\,dy\right|^2.
\end{equation}
This formula is equivalent to formula (\ref{12}) in which
we took $\mu=1$ and added unessential phase factor $\exp
\frac{iX^2}{2\nu}$. Thus the tomogram for ``time moment''
$\nu$ is equivalent to the intensity of the free
propagating signal. In fact, the kernel in (\ref{43}) is
the Green function of a free particle. Since due to
homogeneity the tomogram $w(X,1,\nu)$ is equivalent to
the tomogram $w(X,\mu,\nu)$, while measuring the
intensity of free propagating signal one measures both
the phase and amplitude of the input signal $\psi(y)$. If
one measures the field in optical fiber, the structure of
the output field can be evaluated by measuring the free
propagation of the beam. There is a pecularity in using
formula (\ref{43}). For complete reconstructing the
amplitude, one needs to know the intensity for arbitrary
large values of time (or longitudinal coordinate $z$).
Practically the length or duration can be chosen to fit
appropriate accuracy of the measurement. As an example,
Fig.6 shows the tomographic map and the corresponding
density plot for the case of free propagation of a pulse
whose initial profile is solitonlike with $\beta =1.0$
(also here we have fixed the free parameters as follows:
$V_0 =0$, $E=-1$, and $q_0 =-1$).

\section{Conclusions}
We introduced the tomographic probability distribution
associated with soliton solutions of nonlinear equations.

Nonlinear dynamical equations like nonlinear Schr\"{o}dinger equation
were presented in the form of equation (a nonlinear generalization of
the Fokker--Planck equation) for the standard probability distribution
function.

Specific cases of solitons for a wide family of modified
NLSE were studied in the tomographic representation
explicitly.

The possibility to use tomograms to reconstruct the phase
of linear or nonlinear signals by measuring the signals'
intensities was finally discussed.

\section*{Acknowledgments}
This study was supported by Universit\'{a} "Federico II" di
Napoli, Ufficio Scambi Internazionali -- Programma per la
Mobilit\`{a} di Breve Durata.

This study was also partially supported by the Russian
Foundation for Basic Research under Project
Nos.~00-02-16516 and 01-02-17745.

\end{document}